\documentclass{pasj00}
\begin{document}
\SetRunningHead{C.\ Akizuki and J.\ Fukue}
{Spherical Relativistic Radiation Flows}
\Received{yyyy/mm/dd}
\Accepted{yyyy/mm/dd}
%tex          2005 1016
%referee      200
%editing      200 

\title{Spherical Relativistic Radiation Flows with Variable Eddington Factor}

%%% begin:list of authors
\author{Chizuru \textsc{Akizuki}
\thanks{Present address: Center for Computational Physics, 
University of Tsukuba, Tennoudai 1-1-1, Tsukuba, Ibaraki, 305-8577} 
 and Jun \textsc{Fukue}} %   \thanks{}}
\affil{Astronomical Institute, Osaka Kyoiku University, 
Asahigaoka, Kashiwara, Osaka 582-8582}
\email{j059337@ex.osaka-kyoiku.ac.jp, fukue@cc.osaka-kyoiku.ac.jp}

%\author{B-Firstname \textsc{B-Familyname}}
%\affil{B-Address of Institute}\email{bbbbb@xxx.xxx.xx.xx}
%\and
%\author{C-Firstname {\sc C-Familyname}}
%\affil{C-Address of Institute}\email{ccccc@xxx.xxx.xx.xx}
%%% end:list of authors

%% `\KeyWords{}' always has to be placed before `\maketitle'.
 %Do NOT move this preamble from here!

\maketitle

%\newpage

\begin{abstract}
We solve spherically symmetric radiation flows 
under full special relativity
with the help of a variable Eddington factor $f(\tau, \beta)$,
where $\tau$ is the optical depth and $\beta$
is the flow velocity normalized by the speed of light.
Relativistic radiation hydrodynamics under the moment formalism
has several complex problems, such as a closure relation.
Conventional moment equations closed 
with the traditional Eddington approximation
in the comoving frame have singularity,
beyond which the flow cannot be accelerated.
In order to avoid such a pathological behavior
inherent in the relativistic moment formalism,
we propose a variable Eddington factor,
which depends on the flow velocity as well as the optical depth,
for the case of the sperically symmertic one-dimensional flow.
We then calculate the relaticistic spherical flow
with such variable Eddington factors
to investigate the case that gas is accelerated by radiative force. 
As a result, it is shown that the gas speed reaches around the speed of light by radiation pressure.
%This suggests the possibility that the radiative force could contribute to acceleration of relativistic flows such as astrophysical jets, black hole winds, and gamma-ray bursts.
\end{abstract}

\section{Introduction}

Relativistic outflows from a luminous central object are observed in various active phenomena; e.g., relativistic jets and winds from microquasars (Mirabel, Rodr\'\i guez 1999; Fender et al. 2004), such as SS~433, GRS~1915$+$105, GRO~J1655$-$40, jets in active galactic nuclei, such as 3C~273 and gamma-ray bursts (M\'esz\'aros 2002).
Several mechanisms were proposed to explain these relativistic outflows, including hydrodynamical, radiative, and magnetic drives.
When the luminosity highly exceeds the Eddington one, the relativistic outflow seems to be driven by radiation pressure of the central object.

So far relativistic outflows or winds driven by radiation pressure in the spherically symmetric case have been studied by several researchers (Castor 1972; Ruggles, Bath 1979; Mihalas 1980; Quinn, Paczy\'nski 1985; Turolla et al. 1986; Paczy\'nski 1990; King, Pounds 2003).
As for numerical calculation, the radiation transfer has been sovled in two or three dimensions by a Newtonian treatment, but has not been resolved sufficiently for a highly relativistic case yet.  

On the other hand, under the traditional moment formalism, the relativistic outflows driven by radiation pressure have pathological behavior (e.g., Turolla, Nobili 1988; Nobili et al. 1991; Turolla et al. 1995; Dullemond 1999; Fukue 2005).
That is to say, moment equations for relativistic radiation transfer can have unphysical critical points.
For example, in one-dimensional relativistic radiation flow using the Eddington approximation in the comoving frame, where the moment equations are truncated at the second order, the singularity appears when the flow velocity becomes $c/\sqrt{3}$.
This is understood as follows (Turolla, Nobili 1988; Nobili et al. 1991; Dullemond 1999; Fukue 2006).
The radiative diffusion may become {\it anisotropic} even in the comoving frame of the gas as a result of what the velocity gradient becomes very large in the direction of the flow when the gaseous flow is radiatively accelerated up to the relativistic regime.
Hence, in a flow that is accelerated from subrelativistic to relativistic regimes, the Eddington factor should be different from $1/3$ even in the optically thick diffusion limit.

As already stressed in the literature (e.g., Nobili et al. 1991), the location of the critical point in the moment equations depends on the choice of a closure relation, and with a suitable choice of the closure relation, the critical point may disappear.
For example, Nobili et al. (1991) adopted a variable Eddington factor, which depends on the optical depth.
However, for the present transfer flow, the critical condition, where the denominator of moment equations vanishes,
contains the flow velocity (e.g., Nobili et al. 1991; Fukue 2006).
Hence, as a natural extension, in the present study we have proposed a variable Eddington factor which depends on the ``flow velocity'' as well as the optical depth.
By adopting such a velocity-dependent Eddington factor, we intend to send the critical point away toward the edge of the speed of light.

%In order to avoid the pathological behavior of such a relativistic regime,
%for a plane-parallel case, Fukue (2006) proposed
%a {\it velocity-dependent Eddington factor},
%which depends on the flow velocity $v$:
%\begin{equation}
%    f(\beta) = \frac{1+2\beta}{3},
%\end{equation}
%where $\beta=v/c$.
%Using such a velocity-dependent variable Eddington factor,
%Fukue (2006) obtained the relativistic radiative flow
%accelerated up to the speed of light
%in the plane-parallel case such as a vertical flow from a luminous disk
%(see also Fukue, Akizuki 2006).

In this paper
we propose a velocity-dependent Eddington factor $f(\tau, \beta)$
for a spherically symmetric case, and
solve the fully special relativistic spherical outflows
driven by radiation pressure using such $f(\tau, \beta)$.

In the next section
we propose a variable Eddington factor
for spherical relativistic radiative flows.
In section 3
we describe the basic equations for relativistic outflow
driven by radiation pressure under the spherical symmetry.
In section 4
we show our numerical results of the radiative flow.
The final section is devoted to concluding remarks.

%%%%%%%%%%%%%%%%%%%%%%%%%%%%%%%%%%%%%%%%%%

\section{Variable Eddington Facor}

In this section
we propose and explain
an {\it optical depth and velocity}-dependent variable Eddington factor,
which enables us to treat a problem of relativistic radiation hydrodynamics
in a spherically symmetric case.

\subsection{Traditional Eddington Factor}

We solve the radiation hydrodynamic problem 
semi-analytically using moment equations.
Then, the Eddington approximation is generally used to close moment equations.
The usual Eddington approximation is defined in the comoving frame as
\begin{equation}
P_0=\frac{1}{3}E_0
\end{equation}
where $P_0$ is the radiation pressure and $E_0$ is the energy density both measured in the comoving frame.
This usual Eddington approximation is axiomatic when the radiation field is isotropic.
Such a situation can be satisfied, when the atmosphere is sufficiently optically thick, or when the gas is optically thin with the uniform radiation field.
As is well-known, however, this usual Eddington approximation does not hold when the radiation field becomes {\it anisotropic} in such a case that there is a transition from optically thick to thin states.

When we examine the gas accelerated up to the relativistic speed by radiation pressure, as a clue of mechanism for jets in active galactic nuclei and microquasars, it is necessary to consider a sudden change of optical depth
and the steep velocity gradient.
In order to obtain the terminal speed of the radiatively-driven relativistic outflow, we have to investigate the flow down to the optically thin state.
In addition, the usual Eddington approximation would be violated,
when the gas is accelerated to the relativistic speed
with steep velocity gradient.
%We have to investigate it to aoptically thin state to know terminal velocity of gas.
%In addtionm general Eddington approximation does not consist in the condition
%where gas is accelerated to around speed of light at a stretch.
%Because the gas in comoving frame become under the influence 
%of aberration or blue shift as opposit to the gas in rest frame.
This is because the radiation field may become anisotropic,
even in the comoving frame, due to relativistic aberration and redshift.

\subsection{Optical-Depth Dependent Factor}

In this subsection, we discuss about a better way of dealing with radiation field which is anisotropic.
When there is a transition from optically thick to thin regimes,
for a spherically symmetric case
Tamazawa et al. (1975) set the Eddington approximation as
\begin{equation}
P_0=fE_0
\label{factor0}
\end{equation}
where $f$ is a variable Eddington factor, and they proposed the relation that satisfied the physical condition from optically thick to thin regimes by
\begin{equation}
f(\tau)=\frac{1+\tau}{1+3\tau},
\label{tamazawa}
\end{equation}
where $\tau$ is the optical depth.

This factor becomes 1/3 in an optically thick region while becomes unity in an optically thin region.
This is understood as follows.
The photon mean-free path $\ell$ is on the order of
\begin{equation}
   \ell \sim 1/(\kappa\rho)
\end{equation}
where $\kappa$ is the opacity and $\rho$ is the gas density.
When the gas density is large and the medium is sufficiently thick,
the mean free path becomes small and the radiation field
is locally seen to be isotropic.
While, around the surface of the atmosphere or
in a spherically expanding flow,
the gas density becomes small and the mean free path lengthens
more and more toward the direction of the density gradient;
then the radiation field becomes locally seen to be anisotropic.
In such a transition region,
the relation between the radiation pressure and radiation energy would change in each direction. 
When the optical depth becomes 0, for an outward direction, the radiation pressure is equal to the radiation energy.
However, in the case of relativistic outflow, analytic method can not be calculated until the speed of light with even this factor due to the singularity.

\subsection{Velocity Dependent Factor}

Next, we consider the case where the gas interacting with photon is accelerated to the relativistic speed.
When there is a large velocity gradient, the photon mean-free path becomes longer than that without the velocity gradient.
In such a case, the usual Eddington approximation would be violated again.
For instant, in the relativistic flow with a velocity gradient
$dv/dr$, where $v$ is the flow velocity and $r$ the radius,
the velocity increase at a distance of the mean free path $\ell$ becomes
\begin{equation}
   \Delta v= \ell \frac{dv}{dr} = \frac{1}{\kappa\rho}\frac{dv}{dr}
          \sim \frac{dv}{d\tau}.
\end{equation}
In order for the radiation fields to be isotropic in the comoving frame,
this velocity increase should be sufficiently smaller than
the speed of light; $dv/d\tau \sim v/\tau \ll c$.
If the velocity difference becomes very large when the velocity itself is very high and/or the optical depth is small,
the usual Eddington approximation in the comoving frame would be violated. 
Such a situation can occur for a relativistic outflow.
If the velocity difference is large at a distance of the mean free path, the relativistic effect, such as a Doppler effect and aberration, becomes important, and the radiation field is seen to be anisotropic.

For a relativistic flow with a velocity gradient,
a velocity-dependent variable Eddington factor was proposed (Fukue 2006):
\begin{equation}
f(\beta)=\frac{1}{3}+\frac{2}{3}\beta,
\end{equation}
where $\beta=v/c$.
This factor is applied to the plane-parallel case.
As a good news by the usage of this velocity-dependent factor,
we can avoid  critical points that always appear in the moment equations under special relativity.
In this point, it is indicated that the velocity-dependent variable Eddington factor in the relativistic flow could be reasonable mathematically as well as physically.
However, in the spherical case we also have to consider about the effect of the optical depth against to the plane-parallel case which does not include the effect of the optical depth through velocity of gas.

%\subsection{Eddington Factor Depending both on an Optical Depth and Velocity}
\subsection{Optical-Depth and Velocity Dependent Factor}

Now, we consider the case of a relativistic spherical flow.
The Eddington factor depends on the optical depth for a spherical atmosphere, while it depends on the flow velocity for a relativistic flow.
In the spherically symmetric relativistic flow,
there exist a dilution effect due to a spherical expansion and
that due to a relativistic expansion.
Hence, we suppose that the Eddington factor could depend on both the optical depth {\it and} the flow velocity.
The minimum requirements for such a variable Eddington factor are
(i) it approaches 1/3 in a sufficiently thick, low velocity regime,
(ii) it becomes unity in an optically thin regime, and
(iii) it does also become unity in the relativistic regime at a speed
on the order of the speed of light.
Additional conditions are
(iv) it reduces to the factor of Tamazawa et al. (1975) in a static limit, and
(v) it is simple.

Although there may be many possible factors,
in the present paper we propose the following one,
\begin{equation}
f(\tau,\beta)=\frac{\gamma(1+\beta)+\tau}{\gamma(1+\beta)+3\tau},
\label{factor}
\end{equation}
where $\tau$ is the optical depth, 
$\beta$ is the normalized flow speed ($\beta=v/c$),
and $\gamma$ is the Lorentz factor [$\gamma=1/\sqrt{1-(v/c)^2}$].
This form was born as follows.
It is shown that the mean free path $\ell$ of photons in the inertial frame lengthens than that $\ell_0$ in the comoving frame by a relativistic effect (Abramowicz et al. 1991) as
$\ell = \ell_0/[\gamma (1-\beta \cos\theta)] = \ell_0 \gamma (1+\beta)$.
By considering this, we replaced the optical depth of Tamazawa et al (1975)
by $\tau/[\gamma (1+\beta)]$ for the outward moving flow.
%This form is constructed as follows.
%In the comoving frame,
%we first suppose the variable Eddington factor
%by Tamazawa et al. (1975)
%with an optical depth $\tau_0$ in the comoving frame.
%On the other hand, in the relativistic flow,
%the apparent optical depth depends 
%on the flow velocity (Abramowicz et al. 1991).
%That is,
%the apparent optical depth towards the downstream
%should be $\tau/[\gamma(1+\beta)]$,
%where $\tau$ is the optical depth in the inertial frame.
%Hence, we replace the optical depth $\tau_0$
%by $\tau/[\gamma(1+\beta)]$ to yield the above relation.

Figure 1 shows the behavior 
of the present variable Eddington factor (\ref{factor}).
A dashed curve is the varialble Eddington factor by Tamazawa et al. (1975),
while other curves are the present case for several values of the flow speed.
As is seen in figure 1,
the present variable Eddington factor
becomes unity as the flow speed approaches the speed of light.

%In addition, another variable Eddington factor was given in Fukue(2006) as
%\begin{equation}
%f(\tau,\beta)=\frac{1}{3}+\frac{2}{3}\frac{1+(\tau+1)\beta}{1+\tau+\beta} 
%\end{equation}
%In this paper, we compare two Eddington factors.. 

Using these variable Eddington factors,
we can calculate the spherically symmetric relativistic flow,
continuously from low speed to relativistic regimes.
In the next section, we solve the relativistic moment equations
 with a variable Eddington factor
for the relativistic spherically symmetric case.

\begin{figure}
  \begin{center}
  \FigureFile(80mm,80mm){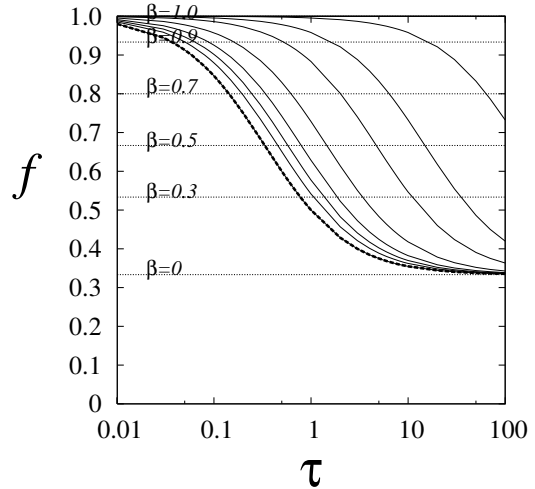}
  \end{center}
  \caption{
Optical depth and velocity dependent variable Eddington factor.
The dashed curve shows an optical depth dependent variable Eddington factor,
$f(\tau)=(1+\tau)/(1+3\tau)$.
The dotted lines denote the velocity dependent variable Eddington factor,
$f(\beta)=1/3+(2/3)\beta$,
where the velocity is $\beta=0.3, 0.5, 0.7, 0.9, 0.99, 0.999, 0.99999$ from bottom to top.
The solid curves represent an optical depth and velocity dependent variable Eddington factor,
$f(\tau,\beta)=\{\gamma(1+\beta)+\tau\}/\{\gamma(1+\beta)+3\tau\}$,
where the velocity is $\beta=0.3, 0.5, 0.7, 0.9, 0.99, 0.999, 0.99999$ from bottom-left to top-right.
}
\end{figure}

\section{Basic Equations}
%The full transfer equation should be solved in order to be comprehended radiation hydrodynamic flows.
%However, the radiation hydrodynamic problem is so difficult that recourse is 
%usually made to approximate solutions of transfer equations 
%(Chandrasekhar 1960; Mihalas 1970; 1986; Mihalas, Mihalas 1984; Kato et al. 1998). 
%We prescribe the gas flow and solve the transfer problems to 
%provide speculations of observation data. 

In this paper, it is treated a simple one-dimensional radiation flow in what follows; i.e., we consider the spherical case in the radial direction.
The radiative energy is transported in the radial direction, and the gas itself also moves in the radial direction by the action of radiation pressure.
For simplicity, the radiation field is sufficiently intense that both the gravitational field, e.g., of the central object, and the gas pressure and the internal heating are ignored in this paper.
%The internal heating is also ignored.
As for the order of the flow velocity $v$, we consider the fully special relativistic regime, where the all terms are retained.
Importance to retain the higher order of velocity is described in Yin and Miller (1995).
Under these assumptions, the radiation hydrodynamic equations for steady radial ($r$) flows are described as follows (Kato et al. 1998; cf. Fukue 2006 for a plane-parallel case).

The continuity equation is
\begin{equation}
  4 \pi r^2 \rho cu = \dot{M} ~(={\rm const.}),
\label{rho1}
\end{equation}
where $\rho$ is the proper gas density, $u$ the radial four velocity, $\dot{M}$ the mass-loss rate, and $c$ the speed of light.
The four velocity $u$ is related to the proper three velocity $v$ by
$u=\gamma v/c$.
% where $\gamma$ is the Lorentz factor,
%$\gamma=\sqrt{1+u^2}=1/\sqrt{1-(v/c)^2}$.

The equation of motion is
\begin{equation}
   c^2u\frac{du}{dr} = \frac{\kappa_{\rm abs}+\kappa_{\rm sca}}{c}
                    \left[ F \gamma (1+2u^2) - c(E+P)\gamma^2 u \right],
\label{u1}
\end{equation}
where $\kappa_{\rm abs}$ and $\kappa_{\rm sca}$ are the absorption and scattering opacities (gray), which relate to bremsstrahlung, photonionization and electron scattering. 
We define them in the comoving frame. 
Moreover, $E$ is the radiation energy density, $F$ the radiative flux, and $P$ the radiation pressure observed in the inertial frame.
In the no-gas pressure approximation and without heating, the energy equation is reduced to a radiative equilibrium relation,
\begin{equation}
   0 =  j - c\kappa_{\rm abs} E \gamma^2 - c\kappa_{\rm abs} P u^2
                  + 2 \kappa_{\rm abs} F \gamma u,
\label{j1}
\end{equation}
where $j$ is the emissivity defined in the comoving frame.
In this equation (\ref{j1}), the third and fourth terms on the right-hand side appear in the relativistic regime.

For radiation fields, the zeroth-moment equation becomes
\begin{eqnarray}
  \frac{1}{r^2} \frac{d}{dr}(r^2 F) &=& \rho \gamma
         \left[ j - c\kappa_{\rm abs} E
             + c\kappa_{\rm sca}(E+P)u^2  \right.
\nonumber
\\
    &&  \left. + \kappa_{\rm abs}Fu/\gamma
               -\kappa_{\rm sca}F ( 1+v^2/c^2 )\gamma u \right].
\label{F1}
\end{eqnarray}
The first-moment equation is
\begin{eqnarray}
   \frac{dP}{dr} &=& - \frac{1}{r}(3P-E)+
           \frac{\rho \gamma}{c} 
         \left[ ju/\gamma - \kappa_{\rm abs} F
                  + c\kappa_{\rm abs}Pu/\gamma \right.
\nonumber
\\
     && \left. -\kappa_{\rm sca}F(1+2u^2)
               +c\kappa_{\rm sca}(E+P)\gamma u \right].
\label{P1}
\end{eqnarray}
Although the first term on the right-hand side of equation (\ref{P1}) disappears for the closure relation such as an Eddington approximation in the optically thick limit, this term remains in the equation because of the modification of the Eddington approximation in this paper.  
It should be noted that this equation (\ref{P1}) is reduced to that by Ruggles and Bath (1979) in the lower approximation of $(v/c)^1$.

Here, in order to close moment equations for radiation fields, we adopt a velocity-dependent or optical depth and velocity-dependent variable Eddington approximation (\ref{factor0}).
%\begin{equation}
%   P_{\rm co} = f(\beta) E_{\rm co}
%\label{close0}
%\end{equation}
%in the comoving frame as the closure relation, where $P_{\rm co}$ and $E_{\rm co}$ 
%are the quantities in the comoving frame.
If we adopt this form (\ref{factor0}) as the closure relation in the comoving frame, the transformed closure relation in the inertial frame is
\begin{equation}
   cP \left( 1 + u^2 - fu^2 \right) = 
   cE \left( f\gamma^2 - u^2 \right) 
   + 2 F \gamma u \left( 1 - f \right),
\label{close}
\end{equation}
or equivalently,
\begin{equation}
   cP \left( 1 - f\beta^2 \right) =
   cE \left( f - \beta^2 \right) + 2F\beta \left( 1 - f \right).
\label{close_beta}
\end{equation}
Above closure relation gives the relation among radiation pressure, energy and flux.
Relations among $E$, $F$, and $P$, which depend on velocity, are important relations, and it is a point on using the modified closure relation. 
%Eddington factors $f(\beta)$ are proposed in the next section.

Eliminating $j$ with the help of equations (\ref{j1}) and using continuity equation (\ref{rho1}), equations (\ref{u1}), (\ref{F1}) and (\ref{P1}) are rearranged as
\begin{eqnarray}
\!\!\!\!\!
   c\dot{M}\frac{du}{dr} &=& 4\pi r^2\rho \frac{\gamma}{c}
                   (\kappa_{\rm abs}+\kappa_{\rm sca})                     
\nonumber
\\ 
  &&             \times \left[ F (1+2u^2) - c(E+P)\gamma u \right],
\label{u2}
\\
\!\!\!\!\!
   \frac{d}{dr} (r^2 F) &=&  r^2 \rho u 
        (\kappa_{\rm abs}+\kappa_{\rm sca})
\nonumber
\\ 
  &&             \times \left[ c(E+P)\gamma u - F (1+2u^2) \right],
\label{F2}
\\
\!\!\!\!\!
   \frac{dP}{dr} &=& -\frac{1}{r}(3P-E)+(\kappa_{\rm abs}+\kappa_{\rm sca})
   \nonumber
\\ 
  &&             \times \rho \frac{\gamma}{c} 
                    \left[ c(E+P)\gamma u - F (1+2u^2) \right].
\label{P2}
\end{eqnarray}
The integration of the sum of equations (\ref{u2}) and (\ref{F2}) yields the energy flux conservation along the flow,
%\begin{equation}
%   c^2 \dot{M} \frac{d\gamma}{dr} + \frac{d}{dr}(4\pi r^2 F) = 0.
%\label{K}
%\end{equation}
%For direction-integral $r$ of equation (\ref{K}), we obtain the relation as follows;
\begin{equation}
c^2 \dot{M}\gamma + L = c^2 \dot{M} + L_0 ~(={\rm const.}), 
\label{L}
\end{equation}
where $L~(=4\pi r^2 F)$ is the luminosity.
The initial conditions are given as $u=0$, $L=L_0$, $P=P_0$, and $r=r_0$ at $\tau=\tau_0$. 
The subscript zero denotes the values at the flow base of $\tau=\tau_0$.
%Besides in this case we do not need to provide final values in boundary condition to be selected with an initial value for suspense parameters.
%Fully special relativistic case however the final values of the radiation fields at the flow top depend on the flow velocity there and the final values at the flow top cannot be analytically expressed by the initial values at the flow base.
On the basis of above the basic equations are the equation of motion (\ref{u2}), the mass flux (\ref{rho1}), the momentum flux (\ref{P2}), the energy flux (\ref{L}) and the closure relation (\ref{close_beta}) at this stage.

Here, we define new variables for convinient calculations: $Q = 4 \pi r^2 cP$ for radiation pressure and $D = 4 \pi r^2 cE$ for radiation energy.
Substituting these variables into equations (\ref{u2}), (\ref{P2}), and (\ref{L}), with the help of equation (\ref{close_beta}), we obtain 
\begin{eqnarray}
c^2 \dot{M} \gamma ^3 \frac{d\beta}{dr} &=& (\kappa_{\rm abs}+\kappa_{\rm sca})\rho \gamma
\nonumber
\\ 
  &&    \times \frac{(f+\beta ^2)L-\beta Q(1+f)}{f-\beta ^2},
\label{u3}
\\
%\!\!\!\!\!
\frac{dQ}{dr}&=&-(\kappa_{\rm abs}+\kappa_{\rm sca}) \rho \gamma \frac{(f+\beta ^2)L-\beta Q(1+f)}{f-\beta ^2}
\nonumber
\\ 
  &&     +\frac{1}{r}\frac{(1-f)(1+\beta ^2)Q-2\beta L(1-f)}{f-\beta ^2},
\label{P3}
\\
%\!\!\!\!\!
\dot{M} c^2 \gamma +L&=&\dot{M} c^2 +L_0.
\label{L2}
\end{eqnarray}
%where $\beta=u/\gamma$.

In addition, we regard the optical depth $\tau$ as
\begin{equation}
    d\tau = - ( \kappa_{\rm abs}+\kappa_{\rm sca} ) \rho dr,
\end{equation} 
and the mass flux (\ref{rho1}), the momentum (\ref{u3}),
the first moment (\ref{P3}), and the energy flux (\ref{L2}) are rewritten as
\begin{eqnarray}
\frac{dr}{d\tau} &=& -\frac{4\pi r^2 c \gamma \beta}{( \kappa_{\rm abs}+\kappa_{\rm sca} )\dot{M}},\\
\label{rho2}
\!\!\!\!\!
c^2 \dot{M} \gamma ^3 \frac{d\beta}{d\tau} &=& -\gamma \frac{(\beta ^2 +f)L-(1+f)\beta Q}{f-\beta ^2},\\
\label{u4}
\!\!\!\!\!
\frac{dQ}{d\tau}&=&-\frac{4\pi rc\gamma\beta}{( \kappa_{\rm abs}+\kappa_{\rm sca} )\dot{M}}\frac{(1-f)[(1+\beta ^2)Q-2\beta L]}{f-\beta ^2}
\nonumber
\\ 
  &&     +\gamma \frac{(f+\beta ^2)L-\beta Q(1+f)}{f-\beta ^2},\\
\label{P4}
\!\!\!\!\!
\dot{M} c^2 \gamma +L&=&\dot{M} c^2 +L_0.
\label{L3}
\end{eqnarray}

In order to transform them into dimentionless forms, the radius $r$ is normarized by the Schwarzschild radius $r_{\rm g}$ ($=2GM/c^2$), the mass-loss rate $\dot{M}$ is normalized by $L_{\rm E}/c^2$, and the pressure $Q$ and luminosity $L$ are normalized by the Eddington luminosity $L_{\rm E}$ [$=4\pi c GM/(\kappa_{\rm abs}+\kappa_{\rm sca})$].
%Then, we transform them in the dimensionless form as follows.  
%The quantities are normalized by
%the Schwarzschild radius $r_{\rm g}$ ($=2GM/c^2$),
%the speed of light,
%and the Eddington luminosity $L_{\rm E}$ of the central object.
%For example,
%the unit of $L$ and $Q$ is $L_{\rm E}$,
%while that of $\dot{M}$ is $L_{\rm E}/c^2$.
They can be rewrirren as
\begin{eqnarray}
\frac{d\hat{r}}{d\tau} &=& -\frac{2 \hat{r}^2 \gamma \beta}{\hat{\dot{M}}},
\label{rho3}
\\
\!\!\!\!\!
\hat{\dot{M}}\gamma ^3 \frac{d\beta}{d\tau} &=& -\gamma \frac{(\beta ^2 +f)\hat{L}-(1+f)\beta \hat{Q}}{f-\beta ^2},
\label{u5}
\\
\!\!\!\!\!
\frac{d\hat{Q}}{d\tau}&=&-\frac{2\hat{r} \gamma\beta}{\hat{\dot{M}}} \frac{(1-f)[(1+\beta ^2)\hat{Q}-2\beta \hat{L}]}{f-\beta ^2}
\nonumber
\\ 
  &&     +\gamma \frac{(f+\beta ^2)\hat{L}-\beta \hat{Q}(1+f)}{f-\beta ^2},
\label{P5}
\\
\!\!\!\!\!
\hat{\dot{M}}\gamma +\hat{L}&=&\hat{\dot{M}} +\hat{L_0}.
\label{L4}
\end{eqnarray}

At this ``first'' normalization stage, we briefly comment the boundary conditions on the present case.
Moment equations are to be solved as a two-point boundary value problem, as is well known.
That is, at the base, flow deep inside the atmosphere, several conditions are imposed on the physical quantities for radiation fields, whereas, at the surface of the atmosphere, some relation generally holds on the radiative moments with or without the external irradiation.
In the present radiative flow, we give the boundary conditions $r_0$, $\beta$($=0$), $Q_0$ (or $P_0$), and $L_0$ at the flow base of the optical depth $\tau_0$.
In addition, there exists some relation for the moment $Q$ and $L$ (Fukue 2006) at the flow top of the optical depth of $\tau=0$.
Then, the mass-loss rate $\hat{\dot{M}}$ should be determined as an eigen value by the boundary condition at the flow top.
Although it indeed be possible we do the ``second'' normalization below.

%In the present treatment, we only consider the radiation field and do not include the gravitational field.
In the present treatment, we only consider the radiation field without gravitational field add up to nothing characterisitic scale expected for mass-loss rate; i.e. mass-loss rate itself can be absorbed in the normalization unit.
%As a result, there is no characteristic scale except for the mass-loss rate,
We further renormalize the variables by $\tilde{r}={\hat{r}}/{\hat{\dot{M}}}$, $\tilde{L}={\hat{L}}/{\hat{\dot{M}}}$, $\tilde{Q}={\hat{Q}}/{\hat{\dot{M}}}$ to yield
\begin{eqnarray}
\frac{d\tilde{r}}{d\tau} &=& -2 \tilde{r}^2 \gamma \beta,
\label{rho4}
\\
\!\!\!\!\!
\gamma ^3 \frac{d\beta}{d\tau} &=& -\gamma \frac{(\beta ^2 +f)\tilde{L}-(1+f)\beta \tilde{Q}}{f-\beta ^2},
\label{u6}
\\
\!\!\!\!\!
\frac{d\tilde{Q}}{d\tau}&=&-2\tilde{r}\gamma\beta \frac{(1-f)[(1+\beta ^2)\tilde{Q}-2\beta \tilde{L}]}{f-\beta ^2}
\nonumber
\\ 
  &&     +\gamma \frac{(f+\beta ^2)\tilde{L}-\beta \tilde{Q}(1+f)}{f-\beta ^2},
\label{P6}
\\
\!\!\!\!\!
\gamma +\tilde{L}&=&1 +\tilde{L_0}.
\label{L5}
\end{eqnarray}
At this ``second'' normalization stage, the mass-loss rate apparently disappears in the basic equations and it seems unnecessary the boundary condition at the flow top to determine the mass-loss rate.
Thus, we solve equations (\ref{rho4})--(\ref{L5}) for a suitable form of variable Eddington factors $f(\tau, \beta)$.

\section{Results and Discussion}

In this section
we briefly show a typical example for the relativistic spherical flow
using the present variable Eddington factor,
and discuss and compare several forms of variable Eddinton factors.

\subsection{Typical Example with Fastest Terminal Velocity}

We first show a typical example of the relativistic spherical flow,
after solving the special relativistic radiation hydrodynamic equations,
using the present proposed factor (\ref{factor}).

Among various combinations of parameters,
we find the case of fastest terminal velocity
for the initial condition at the flow base:
$\tilde{L}_0=1$, $\tilde{Q}_0=0.99$, $\tilde{r}_0=1$,
$\beta_0=0$, and $\tau_0=1$.
In this case, the terminal speed becomes $0.68c$.
The result is shown in figure 2.
As is seen in figure 2,
the gas is accelerated as the luminosity decreases; i.e.,
the radiation energy is converted to the bulk motion
in such a relativistic regime.
The gas is accelerated at around the flow top of $\tilde{r}\sim 5[r_{\rm g}c^2/L_{\rm E}]$,
where the optical depth vanishes.
It is stressed that
there does not appear pathological critical points
inherent in the usual Eddington factor of 1/3.

\begin{figure}  
  \begin{center}
  \FigureFile(80mm,80mm){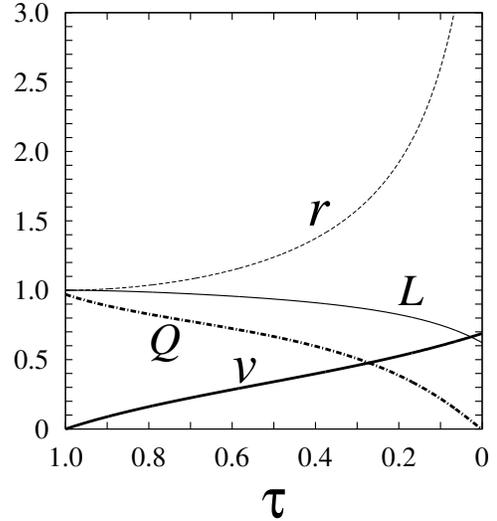}
  \end{center}
\caption{
Flow three velocity $v$ (thick solid curve), 
radiative luminosity $L$ (thin solid curve),
radiation pressure $Q$ (chain-dotted curve),
and radial distance $r$ (doted curve)
as a function of the optical depth $\tau$
for $\tilde{r}_0=1$ at the flow base of $\tau_0=1$.
Other parameters are
$\tilde{L}_0=1$ and $\tilde{Q}_0=0.99$.
}
\end{figure}

Although we can find the relativistic flow
beyond the critical points,
it is difficult to obtain
the solutions with terminal velocities of $\sim c$.
There are several reasons.

The first is the restriction from energy conservation (\ref{L5}).
Because the total energy is constant,
even if all of the radiation energy is converted to the bulk energy,
the terminal Lorentz factor $\gamma_\infty$ is restricted as
\begin{equation}
\gamma_\infty \leq 1 +\tilde{L_0}.
\end{equation}
In the case of figure 2,
the terminal speed is smaller than this absolute limit.

The second reason is the existence of radiation drag.
On the right-hand side of equation (\ref{u6}),
the term of $\tilde{L}$ is the radiative acceleration, while
the term related to $\tilde{Q}$ means the radiation drag force,
which is approximately proportional to the flow speed.
The terminal speed generally becomes high
for large luminosities.
At the same time, however,
the radiation drag force becomes important 
as the flow speed is high.
In the optically thin regime,
radiation drag becomes important.

The third is the dilution (curvature) effect of the spherical flow,
which does not exist in the plane-parallel case (Fukue 2006).
For a simple discussion,
we assume that the flow speed is constant
with the terminal value of $\beta_\infty$ and $\gamma_\infty$.
In such a case, the continuity equation (\ref{rho4})
is integrated from the flow base of $r_0$ to top of $r_\infty$ as
\begin{equation}
   \frac{1}{r_0} - \frac{1}{r_\infty} = 2\gamma_\infty \beta_\infty \tau_0.
\end{equation}
Or, there is a restriction of
\begin{equation}
 \gamma_\infty \beta_\infty < \frac{1}{2r_0 \tau_0}
\end{equation}
for finite $r_0$.
The terminal speed in figure 2
is on the order of this dilution (curvature) limit.

\subsection{Comparison with Various Factors}

In this subsection
we compare the results for various variable Eddington factors.

In figures 3--5
the velocity, the luminosity, and the radius
are shown respectively as a function of the optical depth
for parameters of $L_0=1$, $Q_0=1$, and $r_0=1$.
%The thick solid curve is for the case of $f=1/3$,
%the chain-dotted one for $f(\tau)=(1+\tau)/(1+3\tau)$,
%the dotted one for $f(\beta)=(1+2\beta)/3$,
%the thin solid one for $f(\tau, \beta)
%=[\gamma(1+\beta)+\tau]/[\gamma(1+\beta)+3\tau]$, and
%the dashed one for $f(\tau, \beta)
%=1/3+(2/3)[1+(\tau+1)\beta]/(1+\tau+\beta)$.
The thin solid curve is results for usual Eddington 
factor which is constant.
The chain-dotted curve is the one for the factor of 
Tamazawa et al.(1975) which depends on optical depth.
The dashed curve is the one for the factor is Fukue(2006) 
which depends on velocity.
The thick solid curve is the one for the factor of proposed factor 
in this paper which depends on velocity and optical depth.

As seen in figure 3, the result for usual Eddington factor shows 
an acceleration of the flow is not enough due to the singular or radiation drag. 
On the other hand, in the case of the other factors it can be also shown that the terminal velocity of gas becomes large, even for the same parameters.
In this case there is a difference of around 15\% in each factor.
Although it seems to close these results for Tamazawa's factor and present proposed one, 
the latter is physically acceptable as discussed in section 2.

%As seen in figure 3,
%in the case of the usual factor of $f=1/3$
%the flow is decelerated due to the effect of radiation drag,
%while the flow is accelerated in other cases.
%Although the difference among other cases,
%the energy conversion efficiency is the best for the case of
%the factor $f(\tau, \beta)
%=1/3+(2/3)[1+(\tau+1)\beta]/(1+\tau+\beta)$,
%while it is the lowest for the case of the factor
%$f(\beta)=(1+2\beta)/3$.
%There is little difference between the case of
%$f(\tau, \beta)
%=[\gamma(1+\beta)+\tau]/[\gamma(1+\beta)+3\tau]$
%and the case of
%$f(\tau)=(1+\tau)/(1+3\tau)$
%in this example,
%the former is physically acceptable as discussed in section 2.

As seen in figure 4,
the luminosity is converted to the bulk motion.
That is, the luminosity decrease for the usual factor is small,
whereas that for other factors is large
up to 10\% $\sim$ 15\%.
Such a luminosity change may be a clue to
discriminate the various Eddington factors.

Finally, figure 5 shows the radius change of the expanding photosphere.
%As seen in figure 5,
The terminal speed is large
as the accelerating distance becomes large.
In the present example,
the radius is at most $\tilde{r} \sim 4[r_{\rm g}c^2/L_{\rm E}]$.
For large optical depth,
this radius would be large,
and the terminal speed would also become large.
These results show that gas accelerate as a stretch
in the vicinity of the center of compact objects.

\begin{figure}
\begin{center}
  \FigureFile(80mm,80mm){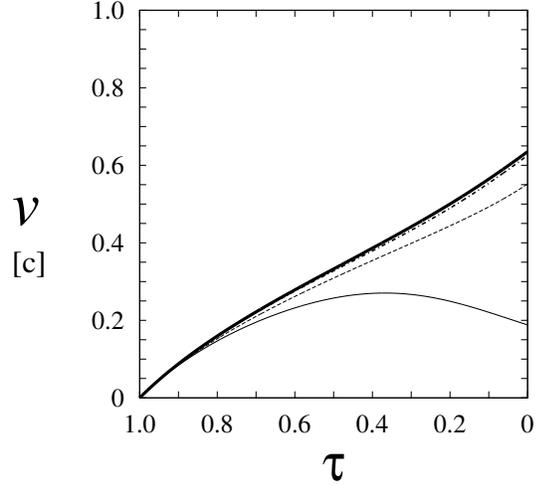}
  \end{center}
  \caption{
Velocity as a function of the optical depth
for parameters of $L_0=1$, $Q_0=1$, and $r_0=1$.
The thin solid curve is for the case of $f=1/3$,
the chain-dotted one for $f(\tau)=(1+\tau)/(1+3\tau)$,
the dashed one for $f(\beta)=(1+2\beta)/3$,
the thick solid one for $f(\tau, \beta)
=\{\gamma(1+\beta)+\tau\}/\{\gamma(1+\beta)+3\tau\}$.
}
\end{figure}
\begin{figure}
\begin{center}
  \FigureFile(80mm,80mm){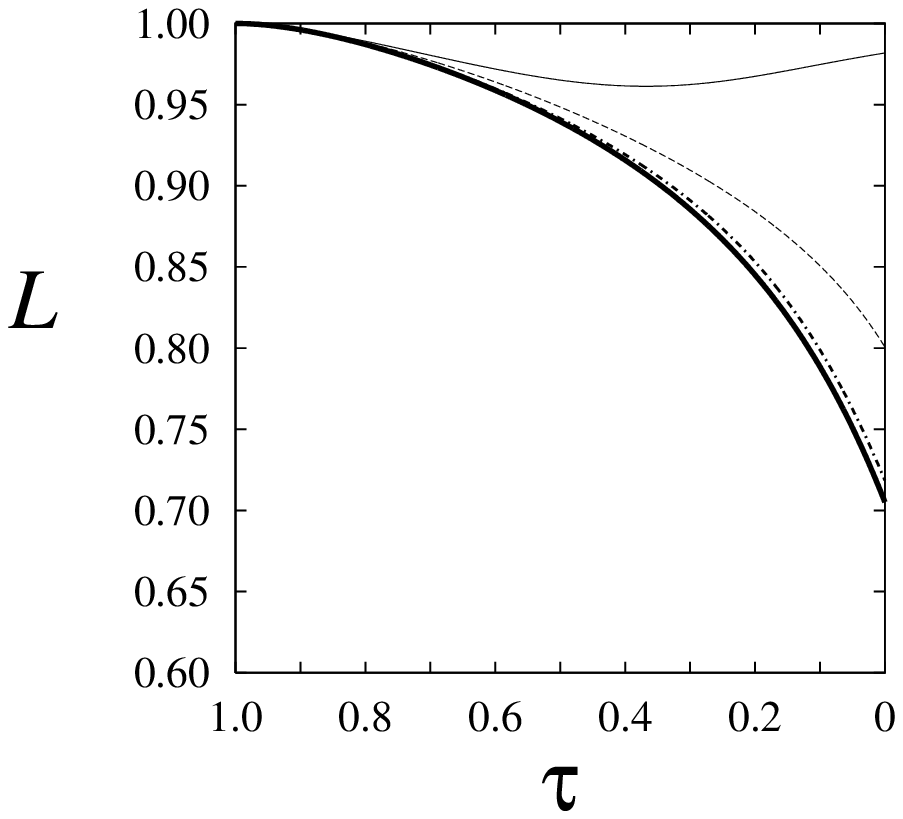}
  \end{center}
  \caption{
Luminosity as a function of the optical depth
for parameters of $L_0=1$, $Q_0=1$, and $r_0=1$.
The thin solid curve is for the case of $f=1/3$,
the chain-dotted one for $f(\tau)=(1+\tau)/(1+3\tau)$,
the dashed one for $f(\beta)=(1+2\beta)/3$,
the thick solid one for $f(\tau, \beta)
=\{\gamma(1+\beta)+\tau\}/\{\gamma(1+\beta)+3\tau\}$.
}
\end{figure}
\begin{figure}
\begin{center}
  \FigureFile(80mm,80mm){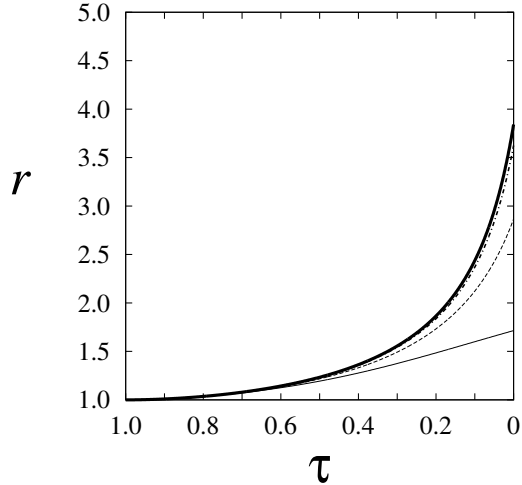}
  \end{center}
  \caption{
Radius as a function of the optical depth
for parameters of $L_0=1$, $Q_0=1$, and $r_0=1$.
The thin solid curve is for the case of $f=1/3$,
the chain-dotted one for $f(\tau)=(1+\tau)/(1+3\tau)$,
the dashed one for $f(\beta)=(1+2\beta)/3$,
the thick solid one for $f(\tau, \beta)
=\{\gamma(1+\beta)+\tau\}/\{\gamma(1+\beta)+3\tau\}$.
}
\end{figure}

%%%%%%%%%%%%%  CONCLUDING REMARKS  %%%%%%%%%%%%%%%%%%%%%%%%%%%%
\section{Concluding Remarks}

In the present paper, we examine the relativistic radiation flow in the spherically symmetric case with the velocity- and optical depth- dependent variable Eddington factors within the framework of special relativity.
We showed that
in the relativisic spherical flow
the Eddington factor is no longer constant,
but depends on the velocity as well as the optical depth.
In particular,
when the gas is accelerated up to the relativistic speed,
there exists a strong velocity gradient, and
the velocity dependence of the Eddington factor becomes important.
In addition, such a variable factor can avoid
the pathological singularity in the moment equations.
We emphasize that we should use such a generalized Eddington factor
to treat the relativistic radiation hydrodynamics
under the moment formalism.

We can find several solutions for the relativistic spherical flow.
The results, however, are slightly different
for the Eddington factor adopted.
In order to determine the precise form of the Eddington factor,
we must solve the relativistic transfer equation rigorously.
However,
the functional form of the variable Eddington factor
is usuful for the study
of the relativistic jets, black-hole winds,
and the gamma-ray bursts.

It should be commented on the current works on the related topics.
Current works are divided mainly 
into two categories, as refered in the introduction.
One type solved the relativistic radiation hydrodynamical equations
under the diffusion approximation
(e.g., Ruggles, Bath 1979; Quinn, Paczy\'nski 1985;
Paczy\'nski, Pr\'oszy\'nski 1986; Turolla et al. 1986; 
Paczy\'nski 1990; Nobili et al. 1994).
In these current works the flow is restricted in the subrelativistic
region on the order of $\sim 0.1~c$.
However,
the diffusion approximatioon may be valid 
only in the sufficiently optically thick regime,
and further, there is no justification that
the diffusion approximation can be used in the relativistic regime,
since there exists a causality problem.
Another type examined the pathological behavior 
of the traditional moment formalism in the relativistic regime
(e.g., Turolla, Nobili 1988; Nobili et al. 1991; Turolla et al. 1995; 
Dullemond 1999; Fukue 2005),
which is one of the motivation of the present study.
However, there is no proposal to use a variable Eddington factor,
which depends on the flow velocity as well as the optical depth,
in order to solve the moment equations in the relativistic regime
in the spherically symmetric case.
We thus tried to solve the relativistic moment equations
with an approximate form of the variable Eddington factor.

In this paper, we considered only the one-dimensinal case without gravity
under special relativity.
In order to clarify the physics of relativistic jets around a black hole,
we must treat the problem within the framework of general relativity.
Such a case is a next work.

\vspace*{1pc}

This work has been supported in part
by a Grant-in-Aid for Scientific Research (18540240 JF) 
of the Ministry of Education, Culture, Sports, Science and Technology.

%\noindent

\end{document}